\documentstyle[11pt,aaspp4,epsf]{article}
\received{}
\accepted{}
\lefthead{Lubow, Seibert, and Artymowicz}
\righthead{Disk Accretion Onto High-Mass Planets}

\newcommand {\apgt} {\ {\raise-.5ex\hbox{$\buildrel>\over\sim$}}\ }
\newcommand {\aplt} {\ {\raise-.5ex\hbox{$\buildrel<\over\sim$}}\ }

\begin{document}

\title{Disk Accretion Onto High-Mass Planets}

\author{S. H. Lubow\altaffilmark{1},
M. Seibert\altaffilmark{1,2}, and P. Artymowicz\altaffilmark{3}} 
\altaffiltext{1}
{Space Telescope Science Institute, 3700 San Martin Drive, Baltimore, MD
21218 lubow@stsci.edu seibert@stsci.edu}
\altaffiltext{2}
{Johns Hopkins University, Department of Physics and Astronomy,
34th and Charles Street, Baltimore, MD 21218}
\altaffiltext{3}{ Stockholm Observatory, 
S-133 36 Saltsjobaden, Sweden pawel@astro.su.se}

\begin{abstract} 

We analyze the nonlinear, two-dimensional response of a gaseous, viscous protoplanetary disk
to the presence of a planet of one Jupiter mass ($1 M_J$)
and greater that orbits a $1 M_{\odot}$ star
by using the ZEUS hydrodynamics code
 with high resolution near
the planet's Roche lobe.
The planet is assumed to be in a circular orbit about the central
star and is not allowed to migrate. 
A gap is formed about the orbit of the planet, but
there is a nonaxisymmetric flow through the gap and onto the planet.
The gap partitions the disk into an inner (outer) disk that extends inside
(outside) the planet's orbit.
For a 1 $M_J$ planet and typical disk parameters,
the accretion through the gap 
onto the planet is highly efficient. That is, the rate
is comparable to the accretion rate toward the central star
 that would occur in the absence
of the planet (at the location of the planet). For typical
disk parameters, the mass
doubling timescale is  less than $10^5$ years, considerably
shorter than the disk lifetime.
Following  shocks near the L1 and L2 Lagrange points, disk material
enters the Roche lobe in the form of
two gas streams.
Shocks occur within the Roche lobe as the gas streams collide,
and shocks lead to rapid inflow towards the planet
within much of planet's Roche lobe. 
Shocks also propagate in the inner and outer disks that orbit  the star.
For  higher mass planets (of order 6 $M_J$), the flow rate onto the planet
is considerably
reduced, which suggests an upper mass limit to planets
in the range of 10 $M_J$. This rate reduction is related to the fact
that the gap width increases relative to the
Roche (Hill sphere) radius with increasing planetary
mass. The flow in the gap affects planetary migration.
For the $1 M_J$ planet case, mass can penetrate from the outer disk to the inner disk, so that
the inner disk is not depleted.
 The results suggest that most of the mass in gas giant planets
is acquired by flows through gaps.

\end{abstract}
 

\section{Introduction}

A gas giant planet is believed to 
form by the accretion of gas
from a protoplanetary disk onto a solid core
(see review by Lissauer 1993).
For typical parameters appropriate to protoplanetary disks,
a planet of mass 1 $M_J$ is expected to open a
gap about the orbit
of the planet (Lin \& Papaloizou 1986, 1993). This result comes
about because the tidal torques exerted on the disk
by the planet (which act to open a gap) are able to 
overcome the viscous torques (which act to close the gap)
(Goldreich \& Tremaine 1978).
This  balance of torques suggests
that for planet masses greater than about  $1 M_J$,
a gap is cleared in the disk.
In this picture, the formation of the gap terminates the
planet's growth.

SPH simulations indicate that
for eccentric binary
star systems, the torque balance argument correctly predicts
the gap locations  (Artymowicz \& Lubow 1994). However, disk material
that surrounds a binary system is sometimes able to penetrate
the gap and accrete onto the binary (Artymowicz \& Lubow 1996, hereafter AL96).
The  flow is highly nonaxisymmetric and takes the
form of two gas streams that are initiated at two corotational
points (analogous to Lagrange points) of the eccentric
potential. Under some circumstances this flow could be quite
efficient. That is, the accretion rate through the gap could be comparable
to the rate that occurs if the binary were replaced by a single
star, so that there is no tidal field to interfere with the flow.
AL96 speculated that a similar process could
occur in the case of flow onto planets.
 Continued accretion
in the presence of a gap may be of importance in understanding
the formation of the
recently discovered high-mass 
(several Jupiter masses) extra-solar planets (e.g., Mayor, Udry, \& Queloz 1998;
Marcy \& Butler 1998).

In this paper, we describe results from numerical simulations
of the interactions of planets with protostellar disks.
The planet is assumed to be in a circular orbit
about the central star. The planet's orbital radius is fixed
in these simulations and  effects of planetary migration
(e.g., Lin, Bodenheimer, \& Richardson 1996, Ward 1997)
 are ignored. Some early work  was carried
out by Miki (1982). This study recognized many of the
global aspects of the flow, but did not address 
the issue of accretion onto the planet.
Studies similar to the one described here have been recently
reported upon by
Bryden et al (1999) and Kley (1999).
The results presented here provide the highest
resolution to date. We are able to determine
properties of the flow within the planet's Roche lobe.

The outline of the paper is as follows. Section 2 describes
the computational procedure. Section 3 describes some tests
of the simulation code. Section 4 provides results
for a $1M_J$ planet, and section 5 provides some results
for higher mass planets. Section 6 discusses
migration. Section 7 contains a comparison
with other studies. Section 8 has the summary and discussion.

\section{Computational procedure}

\subsection{Basic equations}

Protostellar disks are fairly thin (e.g., Burrows et al 1996).
A typical protostellar disk is expected to have a thickness to radius
ratio $H/r \simeq 0.05 - 0.1$. Because of this, the disk
is approximated as two-dimensional by vertically averaging
the three-dimensional fluid equations over height.
Although the vertical averaging retains much of the main
character of the flow, in some regions of space the approximation
is marginally satisfied. In particular, near the Roche lobe of the planet,
the disk thickness is not very small compared
with the radial length scale of the flow (the ratio is about 0.3).

We assume a viscous model for the disk turbulence, with the usual
$\alpha$ disk prescription. The viscous forces are
represented by a Navier-Stokes viscosity force per unit area $\bf{f}$.
The flow is analyzed in the orbital frame that rotates
with the angular speed of the planet $\Omega_p$. In this frame,
the flow achieves a near steady state. We adopt cylindrical coordinates
$(r, \theta)$ with associated flow velocities in the rotating frame
${\bf u} = (u_{r}, u_{\theta})$. The origin of the coordinate system is taken to be the
center of the star. (We ignore the slight center of mass shift
due to the planet.)  The disk self-gravity is ignored.
The equations of motion for the disk of surface
density $\Sigma$ are

\begin{equation}
\label{cons-mass}
\frac{\partial{\Sigma}}{\partial t} + \bigtriangledown (\Sigma {\bf u}) = 0,
\end{equation}

\begin{equation}
\label{rad-force}
\frac{\partial p_r}{\partial t} + 
\bigtriangledown (p_r {\bf u}) =
\Sigma r (u_{\theta}/r + \Omega_p)^2 - \frac{\partial p}{\partial r}
- \Sigma \frac{\partial \Phi}{\partial r} + f_{r},
\end{equation}

and

\begin{equation}
\label{ang-mom}
\frac{\partial j}{\partial t} +
\bigtriangledown (j {\bf u}) =
- \frac{\partial p}{ \partial \theta}
- \Sigma \frac{\partial \Phi}{ \partial \theta} + r f_{\theta},
\end{equation}
where $p_r = \Sigma u_{r}$ is the radial momentum per unit
area, 
 $p$ is the vertically integrated gas pressure, 
$j = \Sigma r (u_{\theta} + \Omega_p r) $ is the gas angular
momentum per unit area,
$\Phi$
is the gravitational potential in the orbit plane due 
to the central star and the planet,
and  $(f_r, f_{\theta})$ is the viscous force per unit area that describes
the effects of disk turbulence.
The gravitational potential $\Phi$ is smoothed in the
neighborhood of the planet with
\begin{equation}
\Phi({\bf r}) = -\frac{G M_{s}}{r} 
       -\frac{G M_{p}}{ [ |{\bf r - r_p}|^2 + r_{sm}^2]^{1/2} },
\end{equation}
where $M_s$ and $M_p$ are the star and planet masses respectively,
$\bf{r_p}$ is the planet location and $r_{sm}$ is the smoothing length
of the planet's potential. Typically, we have chosen the smoothing
length to be 0.2 times the size of the planet's Roche lobe or $r_{sm} =
0.2 a (M_p/(3 M_s))^{1/3}$, where $a$ is the radius of the orbit. 
The smoothing length did not
play an important role in our calculations. We have run models
with a smoothing length of zero and found no important differences
in the general flow properties.

Equations (\ref{cons-mass}), (\ref{rad-force}), and (\ref{ang-mom})
express conservation of mass, the radial momentum, and the 
angular momentum equations, respectively. Equation (\ref{ang-mom}) is written
in terms of the angular momentum $j$, rather than the
azimuthal momentum $p_{\theta}$ in the rotating frame.
The reason is that the $j$ equation provides better
numerical stability (Kley 1998).

The equation of state is taken to be locally isothermal,
 $p \propto \Sigma  T $,
with the temperature expressed as a specified function of radius, T(r).
This equation of state is appropriate for a gas that radiates
internal energy gained by shocks with high efficiency. 
The viscosity force $\bf{f}$ is assumed to be the standard Navier-Stokes
force 
(see Eq 15.3 of Landau \& Lifshitz 1975).
The coefficient of shear viscosity $\mu$ represents the effects of
disk turbulence, while the bulk viscosity coefficient $\zeta$ is set to zero.
The value for the kinematic turbulent
viscosity $\nu = \mu/\Sigma$ is expressed in terms of the usual
$\alpha$ prescription of Shakura and Sunyaev (1973). Namely,
for a disk with local sound speed $c_s$ and thickness $H$,
dimensionless parameter $\alpha$ is defined through
\begin{equation}
\label{alpha}
\nu = \alpha c_s H.
\end{equation}

The use of the $\alpha$ prescription for disks
imposes some simplifications. The actual disk turbulence may
well be magnetic in origin (Balbus and Hawley 1991).
The full effects of MHD turbulence and the associated nonturbulent
magnetic field effects are likely more complex. 
The interaction of waves in the disk with the turbulence
involves additional complexities because the wave forcing
timescale can be shorter than the turnover timescale for the
largest eddies. However, the main
purpose of the disk viscosity is to provide a mass accretion
toward the vicinity of the planet.
The details of the turbulence  might not
be critical for understanding the flow properties near
the planet, where tidal forces from the planet are strong.

Equations (\ref{cons-mass}), (\ref{rad-force}), and (\ref{ang-mom})
are integrated at each time step to provide new values
for $\Sigma$, $p_r$, and $j$, which in turn determine 
the new values of $u_{r}$ and $u_{\theta}$.
The above equations are nondimensionalized so that
the unit of time is the inverse of the planetary orbital frequency 
$\Omega_p$. The unit of distance is the orbital radius of the planet
$a$. The unit of gas density in the disk is $M_d/a^2$, for
a disk of (simulated) mass $M_d$.

\subsection{Numerical method}

We have based our calculation on the ZEUS3D code in two dimensions
(Stone \& Norman 1992).
The code was modified to include a standard Navier-Stokes viscous
force term. The ZEUS code uses operator splitting in which
explicit time derivatives on the left-hand sides of equations 
(\ref{rad-force}) and (\ref{ang-mom}) are determined by 
the terms (called source terms) that appear on the right-hand sides
of their respective equations. Divergence terms on the
left-hand sides of the equations 
(\ref{cons-mass}) - (\ref{ang-mom}) contribute additional changes
in each time step. 

The equations are solved on a spatial grid in the
$r$ and $\theta$ directions. The code allows for variably
spaced gridding, which permits us to obtain higher resolution
in the neighborhood of the planet. The timesteps satisfy the usual
Courant condition for which we have adopted Courant number  0.5.
The code provides an artificial viscosity term, but since we
introduce a Navier-Stokes viscous force, we suppress this
artificial viscosity.
Of course, the code has some intrinsic artificial viscosity
due to the finite gridding. For a uniformly spaced mesh,
the code is formally second order accurate in space and first
order accurate in time. For variably spaced meshes, the code is 
formally first order accurate in space. But, a high level
of accuracy and resolution can be attained by limiting the fractional change
in mesh spacing between adjacent cells to be small, of order one percent.
The van Leer interpolation option was used.

\subsection{Boundary conditions}

The code involves the use of a set of two {\it ghost} zones 
that reside beyond the 
inner and outer boundaries. They are needed to provide spatial
derivatives for quantities in the active zones 
(zones in the simulated region of space) near the boundary.
 The ghost zones were initialized using the same equations
as the active zones. In this way, the velocities in the ghost zones 
of a Keplerian disk retain the Keplerian shear.
The ghost zones get updated at each timestep
to provide a  spatially smooth variation with the active grid zones.
The smoothness is important for the viscous force terms, which depend on the
second derivatives of velocity components.

Boundary conditions need to be specified at the inner
and outer radial boundaries. (The azimuthal boundary
conditions require periodicity in $\theta$.)
For some boundary conditions, highly unstable behavior was found.
If the flow has a steep density
profile near the boundary, then a reflecting or outflow boundary
condition can be applied. For a reflecting boundary condition,
the code sets the radial velocities of the two ghost zones to the sign-reversed
values of radial velocities in the second and third set of active zones
from the radial boundary.
That is, the radial velocities of the ghost zones are reflected
values of the active zones. In addition, the radial velocities in the
active zones closest to the radial boundary are set to zero.
For an outflow boundary condition,
the code sets the radial velocities of the
active zones 
closest to the radial boundary to zero, if the velocity
is directed away from the boundary (inflow). Otherwise, the radial velocity
is left unchanged (for outflow). 

If the density
distribution is fairly flat in radius near the boundary, then an inflow-outflow
boundary condition can be applied. This boundary condition lets
material flow into or out of the active (simulated) region of the system.
The inflow-outflow boundary condition does
not change the velocities in the active zones near the boundaries.
To provide smooth spatial behavior with this boundary condition,
we update the ghost zones with a {\it constant acceleration} scheme.
This is done by applying the changes in physical quantities
of the ghost zones at each timestep by an amount equal to the changes
in nearby active zones. 

\section{Tests of the code}

%NO ARTIFICIAL VISC TERM
We have carried out some tests of the code for accretion disks.
They were performed in the inertial frame without a planet, so that $\Omega_P$
and $M_p$ were taken to be zero in the dynamical equations (\ref{cons-mass}) -
(\ref{ang-mom}). 
In one case, we checked on the stability
of an orbiting disk of initially
constant surface density and pressure. The disk sound speed
was constant in time and space, such that
$H/r$ was set to 0.1 at the radial
midpoint of the simulated region, and viscosity parameter $\alpha$ was set to zero.
The disk was taken to have a ratio of outer to inner radii
equal to 3.  The zoning was 80 by 80, considerably cruder than
our planetary simulations. Inflow-outflow
boundary conditions were applied
at the radial inner and outer boundaries.
The disk was stable, with density fluctuations
less than 1\%  for over
100 orbits, as measured at the disk midpoint.
Although this test may seem trivial, it did reveal several
problems with our early use of the code. 
For example, with a uniformly spaced grid, dynamical instability
was found at the disk inner edge. 
This instability was suppressed
through the introduction of variable mesh spacing or a small
level of viscosity in the Navier-Stokes term $f$, both of which
are used in our planetary simulations.

In another test, we set up a narrow axisymmetric ring of gas that
surrounds a central point mass. The evolution
of the ring was followed numerically and compared with
the analytical solution of a viscous ring (Pringle 1981).
Since the rate of spreading depends on the viscosity, the
test provides a means of checking the 
value of the imposed viscosity parameter $\alpha$
in the Navier-Stokes force $f_{\theta}$.
An 80 by 80 uniform
grid was used, with a ratio of outer to inner radii
equal to 3.
The outflow boundary condition was applied
at the radial inner and outer boundaries.
The density evolution closely followed the analytical solution
(to within a few percent) for several values of viscosity
$\alpha \geq 10^{-4}$. The evolution
was followed over timescales sufficiently long for
substantial profile evolution to occur (40 to 100 orbits at the disk
midpoint).
With the Navier-Stokes term set to zero, the evolution was very
slow. The equivalent level of artificial viscosity
intrinsic to the code was $\alpha \ll 10^{-4}$.

\section{Simulation of a $1M_J$ planet}

\subsection{Boundary and initial conditions}

The planet was assumed to be able to accrete material without
substantial expansion of its radius on the scale of its Roche lobe.
%This condition should be satisfied for gas giant planet masses
%considered in this paper (Lissauer and Bodenheimer xxx).
The planet was fixed on a grid point in the corotating frame at location
$r=1$, $\theta=\pi$. To simulate the accretion of material by the planet,
material that surrounds some region about the planet was nearly fully
 removed from the 
simulation in each timestep. A residual density
of $1.75 \times 10^{-5}$ (in units of $M_d/a^2$)
was retained in these zones, about 0.1\% of the
main disk density near the gap. The removed material was assumed to be accreted
onto the planet, although the mass of the planet was not increased
due to the accretion.  The region of evacuation was chosen
to be the 4 cells that surround the planet.
The evacuation produces a pressure force 
that is directed towards the planet, although this pressure
force is small compared with other dynamical forces.

We imposed
reflecting boundary conditions at the disk inner edge
and inflow-outflow boundary conditions at the disk
outer edge. The disk inner edge was located
at $r = 0.3$ (in units of planet orbital radius $a$),
and the disk outer edge was located at $r=6$.
A variable grid of 252 by 320 was applied, with highest
resolution close to the planet. The grid was uniform
in both $r$ and $\theta$ in a region that included the planet's
Roche lobe.
 With this resolution,
the Roche lobe of the planet contained approximately 
250 cells.

A temperature profile
$T(r) \propto r^{-1}$ was used at all times.
The temperature was normalized such that
$H/r = c_s/(\Omega_p r) = 0.05$ at $r=1$.
The kinematic disk viscosity 
$\nu$ was set to $10^{-5}$ (in dimensionless units
of $a^2 \Omega_p$), independent of radius or time. 
This corresponds to a disk viscosity parameter $\alpha = 4 \times 10^{-3}$
at $r=1$ (see equation (\ref{alpha})).

The initial disk density profile was  chosen
to be axisymmetric and
to follow $\Sigma(r) \propto r^{-1/2}$, together
with a low constant density "gap" imposed near the planet.
The initial gap size and structure was estimated by an approximate
torque balance condition between viscous and tidal torques near the planet.
The gap density was
about $1.75 \times 10^{-4}$ or about 1\% of the disk density just outside the gap.
Computer time was saved by imposing the initial gap,
because a smaller and less violent disk adjustment was 
required to reach a steady state. 
The gap partitions the disk into 
regions interior and exterior to the planet which will be called
the inner and outer circumstellar disks, respectively.

\subsection{Mass accretion rate onto the planet}

In Figure 1 are plotted the azimuthally averaged surface
density profiles at various times for a $1 M_J$ planet. 
The profile achieved a steady state by 100
planetary orbits. The profile indicates the presence
of nonlinear waves in the outer disk. The outer disk was well resolved
and the density variations near the outer boundary were rather smooth.
The inner disk was not resolved as well, and there
is discontinuous behavior of the density at the innermost zones.
 This is partly because
of its more rapidly varying density and velocity with radius.
The inner boundary flow limited the timesteps of the code.

As seen in Figure 2, the planet opens a gap in the disk
and causes waves to propagate in the inner and outer circumstellar disks.
In the outer disk,
the waves damp considerably before reaching the outer boundary.
As seen in Figure 3,
the mass accretion rate onto the planet becomes fairly steady
after about 150 orbits. The accretion rate corresponds
to a gain in planetary mass of $2.7 \times 10^{-5} M_d$
per planetary orbit.
For a typical disk mass of $0.02 M_{\odot}$, this corresponds
to a mass accretion rate of $5.4 \times 10^{-4} M_J$ per orbit
or a mass doubling timescale of about $2 \times 10^{4} yr$ at the orbit of Jupiter.
Since this timescale is much shorter than the disk lifetimes,
estimated as a few million years (Walter et al 1988, Strom et al 1989,
Beckwith \& Sargent 1993), this accretion flow in the presence
of a gap can substantially add mass to the planet.

The accretion rate onto the planet can be compared to the rate
that would occur if the planet were not interfering with the
accretion flow.
The accretion efficiency ${\cal E}$ is then defined
by
\begin{equation}
\label{eff}
{\cal E} = \frac{\dot{M}}{3 \pi \nu \Sigma},
\end{equation}
where $\Sigma$ is a characteristic density value just outside
the gap. In the absence of a planet, the efficiency ${\cal E}$
is unity. For the present case, ${\cal E} \simeq 2$, which suggests that
the flow is quite efficient. Efficiencies greater than unity
are possible because of the gradients imposed by the gap.

\subsection{Flow regions}

Figure 4  provides a closer view of the region near the planet
in Cartesian coordinates after 250 simulated planetary orbits.
The star is located to the right of the figure at $x=0, y=0.$
The left (right) plus sign denotes the outer (inner) Lagrange
point L2 (L1). The Roche lobe containing the L1 point is plotted.
The density image shows spiral arms caused by shocks.
These shocks within the Roche lobe are separate from those
seen in Figure 2 within the circumstellar disk.
The  shocks within the Roche lobe are clearly seen as discontinuities 
in the velocity map of Figure 5.

The streamlines are shown as the dashed lines in Figure 4. Critical streamlines
that separate the distinct flow regions are shown in solid white lines.
The critical streamlines are labeled and sometimes overlap.
For example, streamlines labeled {\it a} and {\it b} in the lower left of Figure 4
overlap with each other, but then bifurcate at the X-point
located near the L2 point. Material in Figure 4 to the left of critical
streamline {\it a} is outer disk material within
the gap (the gap is also seen in Figure 2). Material between streamlines
{\it b} and {\it c} forms a gas stream that accretes onto the planet.
Material below streamline {\it d} is on nonaccreting horseshoe orbits.
Analogous statements can be made about the other streamlines,
which surround the L1 region.  Over an entire orbit about the star, the streamlines
do not close.  So streamlines close to a critical streamline
leaving Figure 4 sometimes return as a streamline on the opposite
side of the critical streamline, as will be discussed later.

\subsubsection{Circumstellar disks }

Gap material associated with the inner and outer disks is {\it locally} deflected away from
the planet by shocks (to the left of streamline {\it a} and the right
of streamline {\it h}).
The local shock deflection of material in the inner and outer
disks is in the direction away from the planet.
Such a deflection is expected on the basis of the theory
of spiral shocks
 (see e.g. Roberts \&
Shu 1972 in the context of galactic shocks). However, this is not
the dominant effect over an entire orbit.
Viscous forces act between shocks to
drive material in the outer disk toward the planet. This finding is not surprising,
because the outer disk is behaving like an accretion disk that feeds material
to the planet. 
Viscous forces are also responsible for driving material from
the inner disk outward toward the planet. Part of this effect
is due to the reflecting inner boundary condition.

\subsubsection{Accretion streams}

The material between streamlines {\it b} and {\it c} and between {\it f} and
{\it g} forms narrow gas streams that penetrate the Roche lobe and
ultimately become accreted by the planet. 
Figure 6 shows the mass flux across dotted line
AD in Figure 4. The mass flux monotonically
decreases from the disk to the gap regions (from point A to point D).
The stream mass flux is peaked near outer bounding streamline
{\it b}  and diminishes
towards inner streamline {\it c}.

As seen in Figures 4 and 5, the material between streamlines {\it b}
and {\it c} undergoes a shock as it approaches the Roche lobe.
This shock is part of the global spiral shock wave visible in Figure 2.
As a result, material in the L2 region
has a very low velocity in the corotating frame of the planet.
In fact, at the X-point the velocity is zero.
The material between streamlines {\it b}
and {\it c} then accelerates as it falls into
the deeper potential within the Roche lobe
and forms the accretion stream.
The stream bears some similarity to those
in Roche lobe overflow of close binary stars (Lubow \& Shu 1975).
 The resulting counterclockwise flow circulation about the planet
is fundamentally
determined by Coriolis deflection.
The
pressure forces  play a more important role here than in the binary
star case,
due to the relatively small size of the Roche lobe.
(Roche lobe
overflow theory predicts that
the region of space where pressure, gravitational,
and inertial forces are comparable, is of
order $c/\Omega_p$ in size. For giant planets,
this region is comparable to the radius of the Roche lobe.)
In addition,
the presence of two gas streams causes additional 
interactions communicated by pressure.

The streams undergo a strong shock
on the opposite side of the Roche lobe
(180 degrees away) from their entrance, due to their
mutual collision (see Figures 4 and 5). Following this shock, the
streams lose a considerable amount of their angular momentum about the planet
and execute highly eccentric orbits.
The material spirals inward towards the planet as a result
of successive shocks. The shock waves about the planet show signs of weakening
close to the planet.  The flow region close to the planet
($\sim 0.2$ of the Roche lobe size) shows some indication
of a more circular Keplerian flow pattern, although the 
resolution in this region is limited.
From this inner region, the material is subsequently
accreted by the planet.

On the basis of the theory of spiral
shocks (Roberts \& Shu 1972), the magnitude of radial flow speed $u_{rp}$, relative
to the planet, induced by the shocks
about the planet is approximately given by 
\begin{equation}
u_{rp} \sim \frac{u_{\perp}^2}{ 2 \pi u},
\end{equation}
where $u_{\perp}$ is the flow velocity perpendicular to the shock
wave front.
 As can be seen from the Figure 4, velocity $u_{\perp} \sim u$.
The above equation indicates that
 $u_{rp} \sim u/(2 \pi)$. In other words, the shock geometry
indicates that the shocks are quite strong and this leads to a fairly
rapid radial inflow. The above equation also indicates why the effects
of the shocks on the inner and outer disk flows are considerably weaker. Namely,
the waves are much more tightly wrapped (see Figure 2),
so that $u_{\perp} \ll u$ and therefore $|u_r| \ll u$.

\subsubsection{Horseshoe orbits}

Also seen in Figure 4,
the flow well within the gap, but outside the Roche lobe, follows horseshoe
orbits (e.g., Dermott and Murray 1981). This region is 
below streamline {\it d} and above streamline {\it e} in Figure 4.
The horseshoe orbits possess a higher value of Jacobi
constant (effective energy) than the inner and outer
disk material. The effect of dissipation
on horseshoe orbits is to drive material away from
the orbit of the planet (out of the horseshoe orbit region).

Figure 7 shows two horseshoe orbits traced by streamlines
at a time of 250 planetary orbits. Both horseshoe orbits originate and terminate
near $x=0$, below $y=-1$. As can be seen from the Figure
the orbits are slightly not closed, due to a small
outward drift.

\subsection{Source of accretion streams}

The stream material originates from the disk edges in the gap,
rather than  material on horseshoe
orbits.  
Figure 8 shows some streamlines for material that flows
onto the planet. Material within the bounding streamlines that surrounds
the left arrow in the Figure orbits about the star, and it
returns to constitute the L2 gas stream that is indicated
by the two right arrows in the Figure. As seen in the Figure,
the contribution to the gas stream comes from the disk material
in the gap, rather than material in the horseshoe orbit region.
At various times, we find fluctuations about this state.
Sometimes a small fraction of the stream material does
originate from the outermost parts of the horseshoe
orbit region, near streamline {\it e} in Figure 4. But this material
is found in turn to originate dominantly from the L1 disk region (to the right
of streamline {\it h}),
and so is only marginally and temporarily associated with the horseshoe
orbit region. Some slight
contribution can come from the decaying
horseshoe orbits as seen in Figure 7
(as suggested by
Bryden et al 1999 and Kley 1999).
As can be seen in Figures 4, 5, and 7, the material
well within the horseshoe orbit region 
(below streamline {\it d} or above streamline {\it e})
that approaches the Roche
lobe is deflected away from the Roche lobe by shocks.

%This ratio can also be estimated by analyzing Figure 4.
	
%Figures 4 and 5
%displays a closer view of the Roche lobe region which
%clearly shows a two-armed spiral shock wave.

\subsection{Low mass inner disk}

The reflecting inner boundary condition applied thus far
did not allow material to accrete onto the central star.
Consequently, an inner disk was always present in the simulations.
The case of an initially low mass inner disk demonstrates effects at the opposite
extreme. We constructed a simulation with the same initial
conditions as before, except that densities in the inner disk
were set to the density value at the gap midpoint 
of $1.75 \times 10^{-4}$.

A mass buildup of the inner disk occurred
at a rate of approximately $5 \times 10^{-5} M_d$ per
planetary period over the 300 simulated orbits. 
This accretion rate is approximately twice that
onto the planet in section 4.2. The efficiency
 of this flow is ${\cal E} \simeq 4$ (see eq (\ref{eff})).
This effect will likely be reduced 
when the planet's
radial migration is taken into account.
 This result suggests
that the inner disk material within the orbit of a $ 1 M_J$ planet will
not be depleted to low density by starvation of material from the outer disk.

Figure 9 shows the flow on the scale of the Roche lobe after 150 simulated
planetary orbits.
In comparison with Figure 4, we see that the L2 accretion stream
occupies a geometrically smaller region. 
It also consists of material from the
outer disk, with no significant contribution from horseshoe orbiting gap
material. Some material from the outer disk sweeps through the
Roche lobe and flows into the newly formed inner
disk. Some material from the inner disk has begun to form
a weak L1 stream of material that flows back onto the planet.

Figure 10 traces a streamline that flows from the outer
to inner disks.
This streamline begins in the outer disk and circulates
about the star in the outer portions of the horseshoe
orbit region. It subsequently joins the inner disk flow.

\subsection{Cause of shocks within the Roche lobe}

Figure 4 shows that the shock waves in the inner and
outer circumstellar disks are quite separate (in azimuthal phase)
from the shock waves within the Roche lobe. There are two likely
causes of the shocks within the Roche lobe. One possible cause is the
resonant generation of a spiral wave by the $m=2$ component
of the tidal forcing provided by the star (see e.g., Savonije, Lin, and Papaloizou
1994 for the case of a binary star system). The other
possibility is direct impact of gap material being
fed into the Roche lobe with material  circulating
within the Roche lobe. We can use Figure 9 to help discriminate between
the two possible causes.
If the dominant cause of the shocks were resonant forcing,
then the two arms would have equal strength in Figure 9.
(We have verified this would be true with this code,
by simulating a circumplanetary disk without
circumstellar disks present.)
On the other hand,
unequal strength arms would result in Figure 9,
 if the shocks in Figure 4 were formed
by the colliding streams. In that case, a one-sided shock structure
would be expected in Figure 9 as the stream impacts with itself 
on the L2 side of the Roche lobe.
The results in Figure 9 do show a two-arm pattern, but
with rather different amplitudes in the arms. 
This suggests that collisions dominate the shock production.
Furthermore, it is evident from the Figures 4 and 5
that the shocks
in the outer region of the Roche lobe are due to a direct impact
from stream material.

%The mass radial inflow about the planet is highly nonaxisymmetric and
%reflects the effects to the two shocks (see Figure 6)

\section{Planets of higher mass}

We have conducted a series of simulations
of the planet accretion from disks with planet
masses from 1 to 6 $M_J$.
For historical reasons, the disk conditions were somewhat different from those
described in the last section. The disk thickness to
radius ratio was taken to be constant in radius
with $H/r = c_s/(\Omega_p r) = 0.05$ throughout.
The kinematic disk viscosity was chosen to be 
$\nu = 10^{-5}$ at $r=1$, corresponding to an
$\alpha = 4 \times 10^{-3}$ at $r=1$.
The model parameters were the same as in the last
section at $r=1$, only their variation in $r$ is different.
A uniform $r-\theta$ grid of 100 by 300 was used. The Roche lobe
flow was not well resolved. The mass accretion rate
was determined by evacuating the entire Roche lobe
in each timestep. Consequently, the absolute values
of the accretion rates are not as well determined as in the
last section. However, the relative accretion rates 
for different planetary masses are likely better determined
than the absolute rates.

Figure 11 shows the accretion rate as a function
of planet mass. There is a clear trend for the
mass accretion rate to drop with increasing
planet mass. This decline is related to the fact
that the gap width increases relative to the
Roche (Hill sphere) radius with increasing planetary
mass. 
For a planet of $6 M_J$, the accretion
rate has dropped substantially. The mass accretion rate
is smaller by about a factor of ten. Therefore,
the timescale to accrete 1 $M_J$ 
from a $0.02 M_{\odot}$ disk increases to several hundred thousand years.
For an even higher mass planet, the flow rate would be less.
So it would appear difficult, but not impossible to reach say $10 M_J$
over the disk lifetime (several million years) under these conditions.

\section{Migration}

The planets modeled in this paper are subject
to radial migration due to their interactions
with the gas. The migration occurs in the presence
of a gap in the disk (Lin \& Papaloizou 1986)
and is sometimes called Type II migration (Ward 1997).
In the case that the planet mass
is small compared to the disk mass, the planet
transmits the disk accretion torque across the gap, and is
otherwise carried inward, along with the accretion flow.
The main limitation of the current study
in determining the effects of migration
is that the planet was not allowed to move radially
over the course of the simulation. For a mobile planet,
the torque imposed by the disk would adjust to provide
migration on the viscous evolution timescale
of the disk.

There are, however, two limiting cases where the migration
effects can be considered based on the results
of this paper. In both cases,
the application is valid because the migration rate is small. In one case,
the planet mass is large compared to the disk mass.
Another case occurs
when the planet is surrounded by only an outer disk.
This situation could arise for example
if a planet migrates to a central hole in the disk
where migration effectively stops, enabling
the so-called hot Jupiters to survive capture
by the central star (Lin et al 1996).

Table 1 shows the migration rates due to the gravitational
back reaction of the gas on the 1 $M_J$ planet
model of section 4. Contributions to migration come
from torques in the 4 regions of space described in the Table:
the inner disk, inner gap, outer gap, and outer disk.
In this terminology, the inner and outer disks refer
to the region outside the gap seen in Figure 2.
The inner and outer gap regions contain all the gas
interior and exterior to the planetary orbit respectively
within the gap.
The third column gives the migration rate expressed as
 $(\dot{a}/a) (2\pi/\Omega_p) (M_s/M_d)$, where
$a$ is the orbital radius of the planet and $M_s$ is the mass
of the central star. This column expresses the fractional change
in the orbital radius per planetary orbit, scaled by the
mass ratio of the star to the disk.
Torques arise from the nonaxisymmetric distribution of material
about the planet, as
seen in Figures 2 and 4. The individual effects
of the inner and outer gaps
are strong, about 5 times bigger than the circumstellar disk
contributions. However, they
nearly cancel by establishing opposing torques on the planet.
Their net contribution is an inward migration that is comparable
to the net circumstellar (nongap) disk contributions, which cancel
less completely between the inner and outer disks.
The overall effect is an inward migration at about
the same rate as is due to the outer disk alone.
Thus, the situation is not very different from
Type I migration (Ward 1997).

Consider now the second case, namely where the
planet's migration is halted near the inner edge
of the disk. We have computed the resulting migration
in such a case. The migration rate corresponds closely to
having contributions only from the outer disk and outer gap 
in Table 1. The result is that the planet has a net outward
migration back into disk, caused by the gap contribution. 
On the other hand, if the planet
is embedded well  within the disk, it will migrate inward,
via Type II migration. This suggests that the planet may
remain in an equilibrium point of no net migration by remaining
within the disk and continuing to accrete matter. 
This equilibrium assumes that gas flows
within the gap maintain the torques expected on the basis
of these calculations. In reality, these flows may be affected
by magnetic fields that are responsible for maintaining the central
hole in the disk by carrying material
onto the central star (Koenigl 1991, Ostriker \& Shu 1995). So further investigations are needed.

\section{Comparison with other results}

As mentioned in the Introduction, Bryden et al (1999) and
Kley (1999) have carried out similar calculations,
but with lower resolution of the Roche lobe region.
Our disk conditions were coordinated with those of Kley,
and show good numerical agreement for the accretion rates.
For the same disk mass, the mass accretion
rates we obtain are about a factor of 2 less than those of Kley
for his model {\it 2q}.
The results of the Bryden et al (1999) study also provide
similar accretion rates. Some differences in accretion
rates may be because of their lower resolution of the Roche region.
The accretion rates are also consistent with preliminary
results using the PPM code reported in Artymowicz, Lubow,
and Kley (1998).

\section{Summary and discussion}

Our simulations show that a planet of mass $1 M_J$ or greater
opens a gap in a protostellar disk for typical disk parameters
(see Figure 2). However, mass can penetrate the gap
and flow onto the planet (see Figures 3 and 4).
For a $1 M_J$ planet, the flow onto the planet occurs
with high efficiency. That is, the accretion rate occurs at about the rate
that material would accrete through the planet's orbit in the
absence of the planet (and its associated gap). These mass accretion
rates are consistent with those found by Bryden et al (1999)
and Kley (1999). This high efficiency was shown by Kley to hold over a wide
range of disk turbulent viscosities, even to much smaller values than considered
here.
 This result means that a significant
fraction of the next $1M_J$ of inflowing disk material exterior to a 
$1 M_J$ planet 
would be accreted by the planet.

The flow near planets follows the main principles that govern the
mass flow though gaps around binaries (AL96),
although there are some differences.
In this picture, mass accretion flows originate at corotation points
and take the form of gas streams. In the eccentric binary case, the
corotation points were provided by a dominant component
of the binary eccentric potential. A  large gap
encompassed the entire binary and two streams were evident in simulations.
The streams represented the highest density material in much of
the gap. 

In the planetary case, the gap is much narrower, due to the
relatively small mass of the planet (see Figure 2). Furthermore, the planet
was taken to reside on a circular orbit which limits
the number of resonances present. 
The corresponding corotation
points in the circular orbit case are the L1 and L2 Lagrange points.
As seen in Figures 4 and 5, the flow into the Roche
lobe occurs in the form of two gas streams that penetrate
at these corotation points. The streams are composed
of material that comes from the disk (rather
than the gap horseshoe material), as seen in Figure 8.
This disk material undergoes a shock and nearly comes to rest
in the corotating frame close to the L1 and L2 points (see Figure 5).
The material then accelerates towards the planet as gas streams.
Each stream undergoes a strong shock
at about 180 degree in azimuth around the planet from the point
of penetration, due to a collision with the alternate stream
(see Figures 4 and 5).

Subsequent shocks within the Roche lobe 
drive material
fairly rapidly toward the planet. The density peaks within
the Roche lobe are dominated by these shocks, which give
the appearance of the inner spiral arms in Figure 4. 
Closer to the planet,
there is evidence for a circular Keplerian disk without
strong shocks.

For higher mass planets, the mass flow onto the planet
weakens considerably (see Figure 11). 
This behavior differs from that found for
eccentric binary star systems with an
assumed higher viscosity, where no
apparent tidal limit to the flow was found (AL96).
On the other hand, some mass flow does occur 
through gaps, which can apparently substantially
increase planetary masses beyond $1 M_J$.
The trend seen in Figure 11 is related to the fact
that the gap width increases relative to the
planetary Roche (Hill sphere) radius with increasing planetary
mass.
Based on the current calculation,
it would appear that tidal truncation might limit planet
growth to roughly $10 M_J$ (see section 5).
Such a result would be consistent with the highest mass
extra-solar planets.
The combined number distribution of planets and spectroscopic
 binary stars as a function
of secondary mass suggests a dip in the range of 7 $M_J$
(Mayor et al 1998) or 10-30 $M_J$
 (Mazeh, Goldberg, \& Latham 1998),
indicative of a mass upper limit to planets and lower limit
to stars in this range.

The circumplanetary disk provides a possible site
 for the formation of planetary satellites. The sense of rotation
of the disk is prograde, due to the influence of Coriolis
forces on the gas streams. Consequently, the accretion
flow provides prograde planetary spins and prograde satellite
orbital motion. The more quiescent flow close to the planet
(see Figure 4) may permit a more favorable region for 
satellite accumulation.

The gas flows can also affect the migration, since they come
close to the planet and exert torques (see section 6).
Another potentially important effect
is  planetary eccentricity. The more massive
extra-solar planets are observed to be eccentric
(see Mazeh, Mayor, \& Latham 1996). In some cases, the
eccentricity may have been present while the planet was surrounded
by a disk, in particular if the eccentricity was generated
by planet-disk interactions (e.g., AL96).
The mass flow rate through a gap surrounding an eccentric planet may well 
be higher than for a corresponding circular orbit planet. An
eccentric planet  opens a larger gap, due to its radial
excursions. More resonances occur in such a case and the
situation may become closer to that expected
for eccentric binary star systems.

\acknowledgments

We gratefully acknowledge support from NASA Grants
NAGW-4156 and NAG5-4310, STScI DDRF, and the STScI
visitors program. We also acknowledge fruitful discussions
during the ITP program on planet formation, organized by Doug Lin,
and during the Isaac Newton Institute for Mathematical Sciences program
on accretion disks. We are grateful to John Wood for providing us with
Cray time at GSFC. We thank the referee, Peter Goldreich,
for comments.

\newpage

\begin{figure*}
\figurenum{1}
\epsscale{1.0}
\plotone{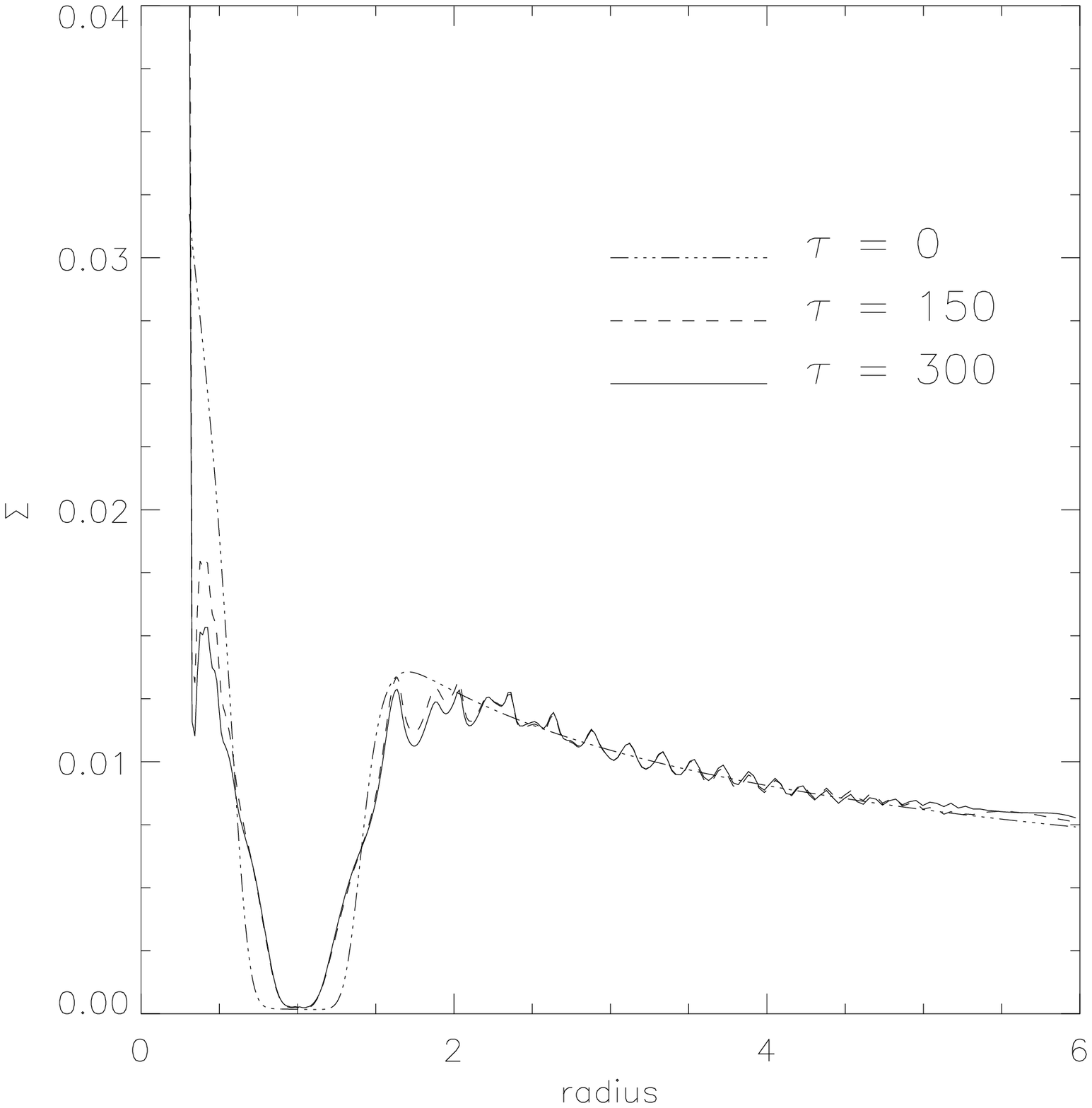}
\caption[fig1.eps]{Azimuthually averaged surface density 
profiles of a disk containing a planet at various times. The density
is normalized by $M_d/a^2$, where $M_d$ is the simulated
disk mass and $a$ is the planetary orbital radius. The horizontal
axis is normalized by $a$.
A $1M_J$ planet resides at $r=1$, and a $1 M_{\odot}$ star
is located at $r=0$. The disk properties
are described in section 4.1. The dashed-dotted line is the initial
density profile, the dashed line is the profile at 150 planetary
orbits, and the solid line is the profile at 300 planetary orbits.
\label{fig1}}
\end{figure*}

\begin{figure*}
\figurenum{2}
\epsscale{1.0}
\plotone{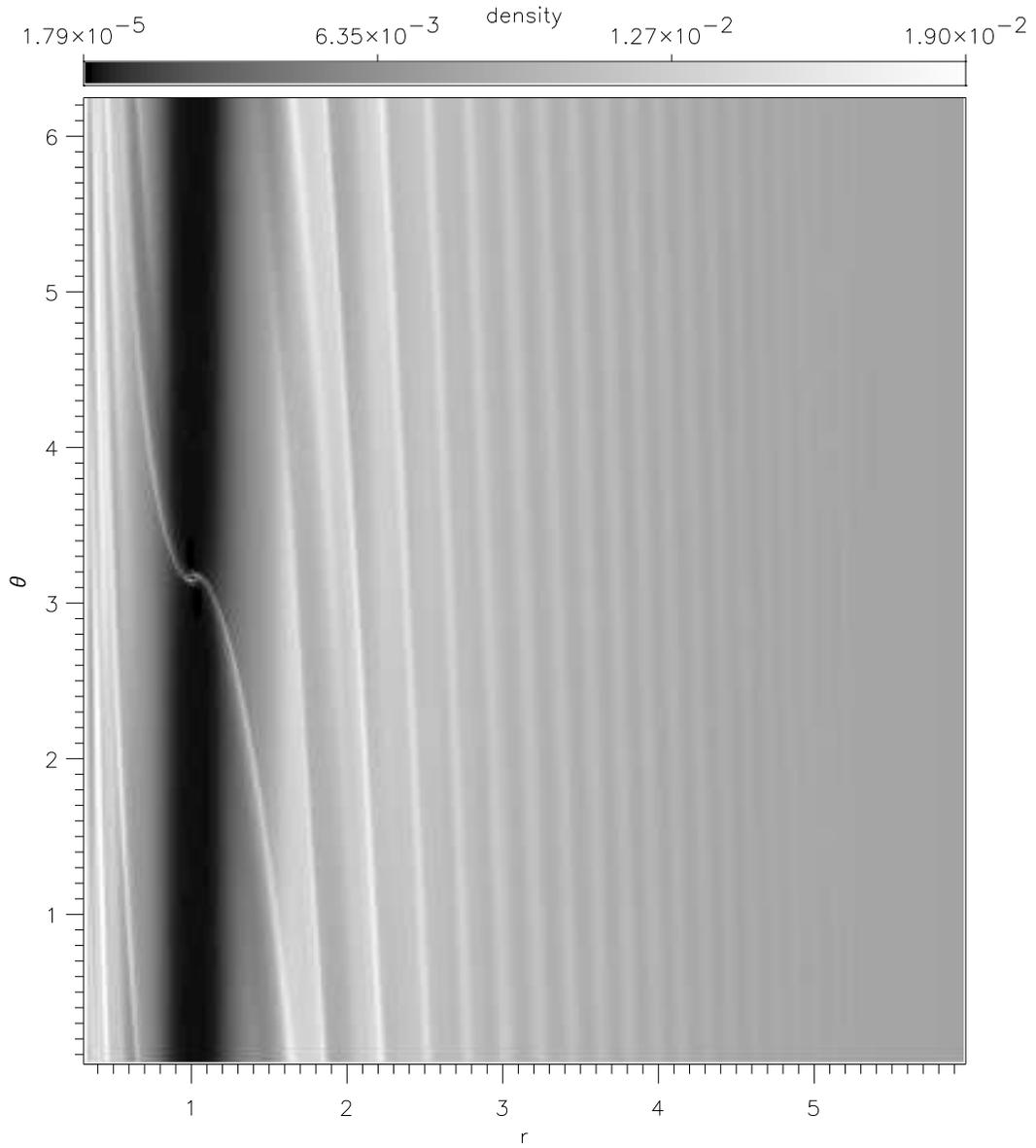}
\caption[fig2.eps]{Density image of the disk
after 300 planetary orbits. The vertical direction
is the angle $\theta$  about the star. The horizontal
direction is the radius $r$ from the center of the star.
The planet is located at $r=1$,
$\theta = \pi$.
\label{fig2}}
\end{figure*}

\begin{figure*}
\figurenum{3}
\epsscale{1.0}
\plotone{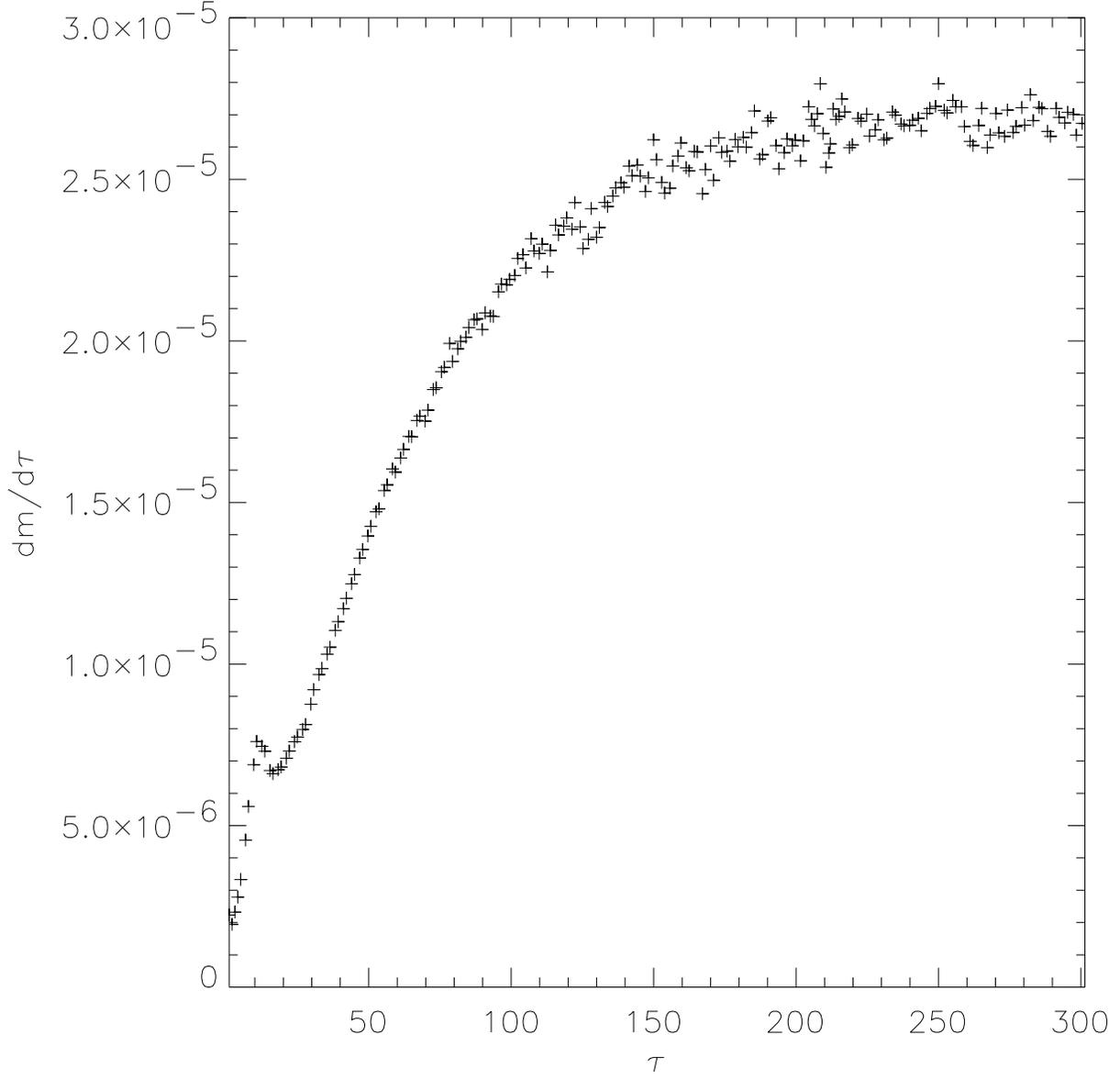}
\caption[fig3.eps]{Mass accretion rate 
onto a $1 M_J$ planet versus time.
The accretion rate is expressed in units of the disk mass $M_d$ per planetary
orbit and the time is in units of planetary periods.
Each point represents the mass accretion rate averaged over 
1 planetary orbit.
\label{fig3}}
\end{figure*}

\begin{figure*}
\figurenum{4}
\epsscale{1.0}
\plotone{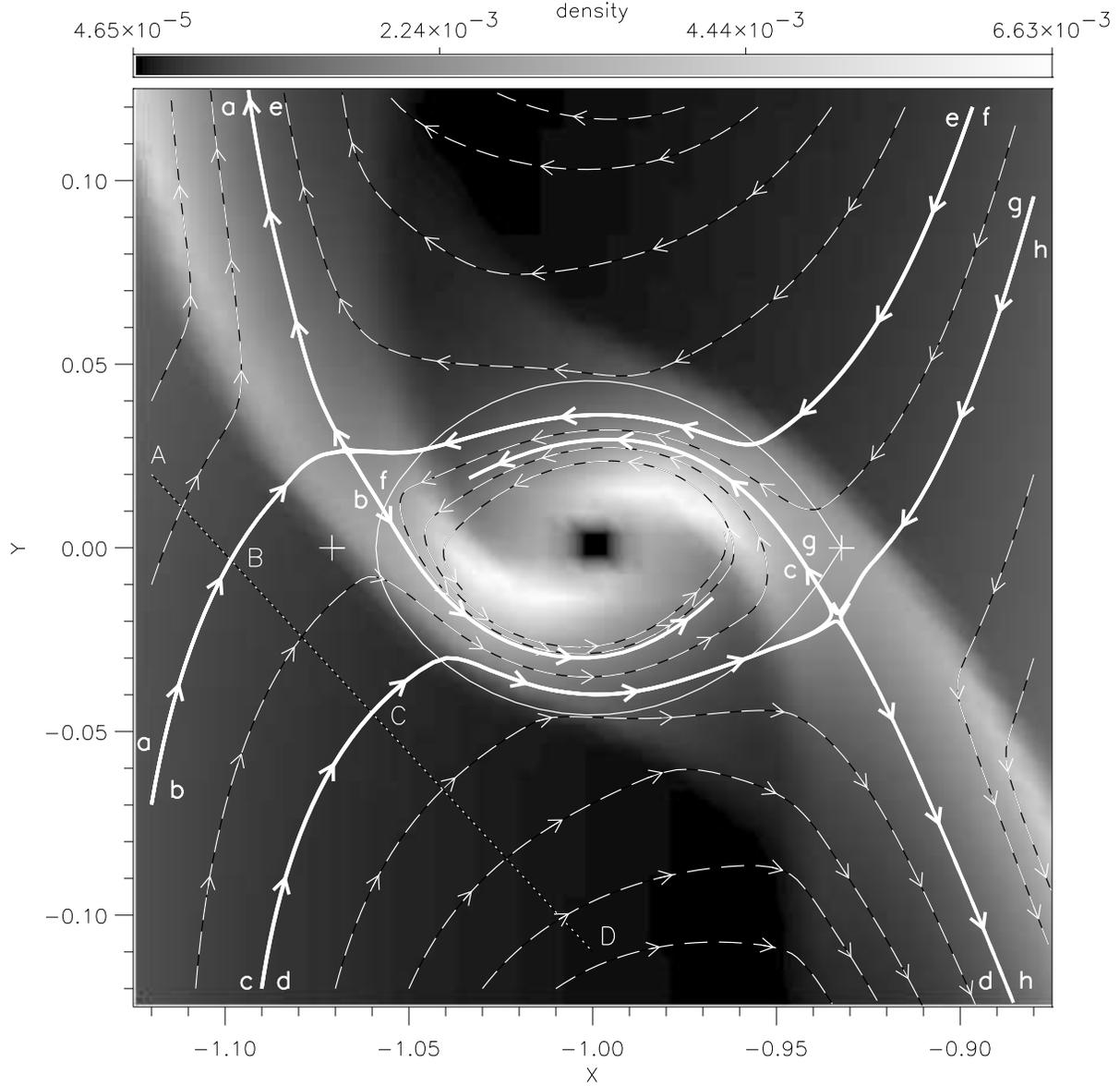}
\caption[fig4.eps]{A high resolution density image
(in Cartesian coordinates) of the Roche lobe region about a $1 M_J$ planet.
The planet is at location x= -1, y=0 and the star is located
at the origin. The left (right) plus sign marks the L2 (L1) point.
Sample streamlines are dashed flow lines. Critical streamlines
that separate distinct flow regions are solid flow lines
(see section 4.3). The dotted line segment AD is used
for Figure 6.
\label{fig4}}
\end{figure*}

\begin{figure*}
\figurenum{5}
\epsscale{1.0}
\plotone{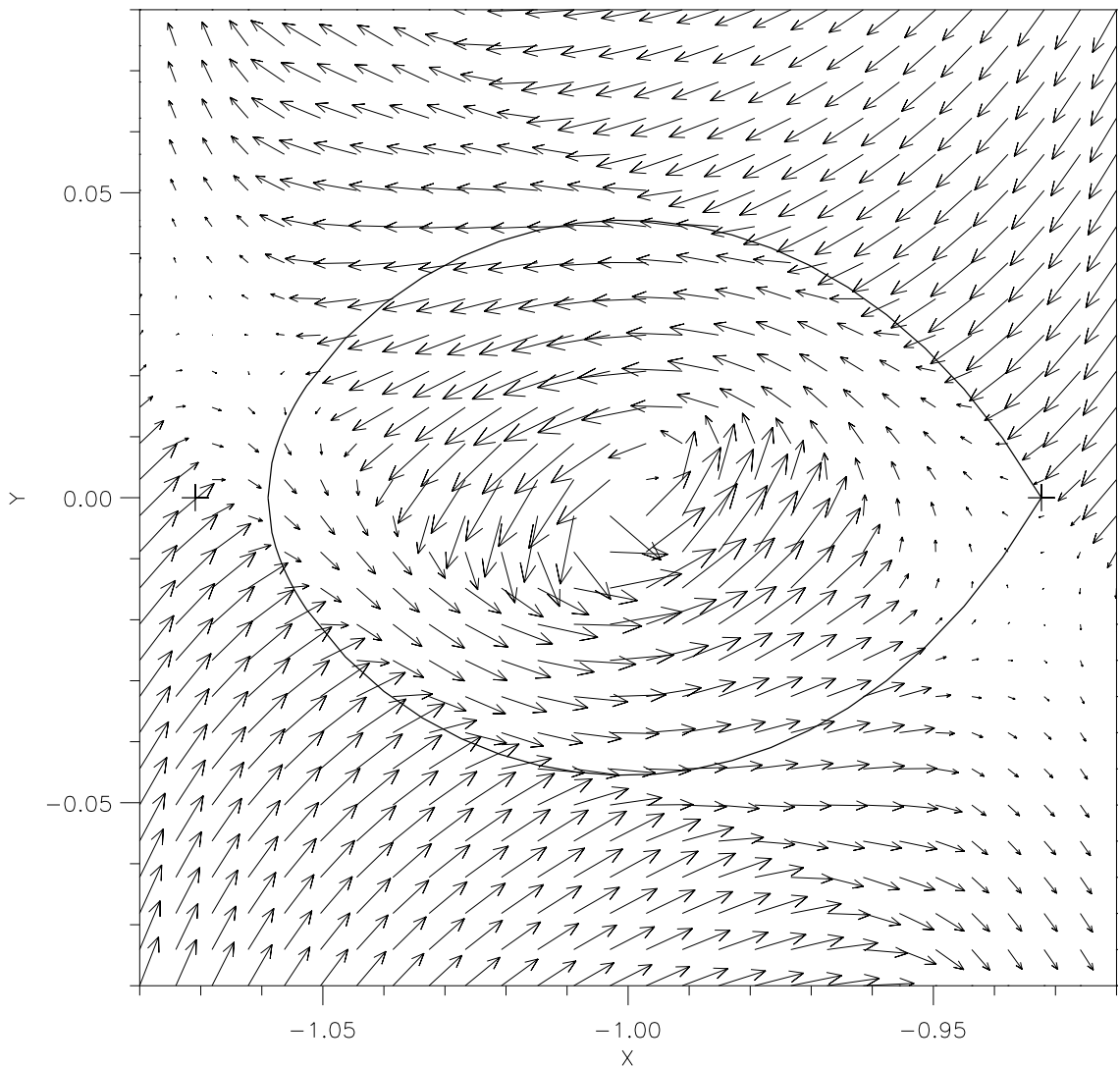}
\caption[fig5.eps]{A high resolution velocity map in the corotating
frame of the planet
(in Cartesian coordinates) of the Roche lobe region about a $1 M_J$ planet.
The planet is at location x= -1, y=0 and the star is located
at the origin. The left (right) plus sign marks the L2 (L1) Lagrange point
(see section 4.3).
\label{fig5}}
\end{figure*}

\begin{figure*}
\figurenum{6}
\epsscale{1.0}
\plotone{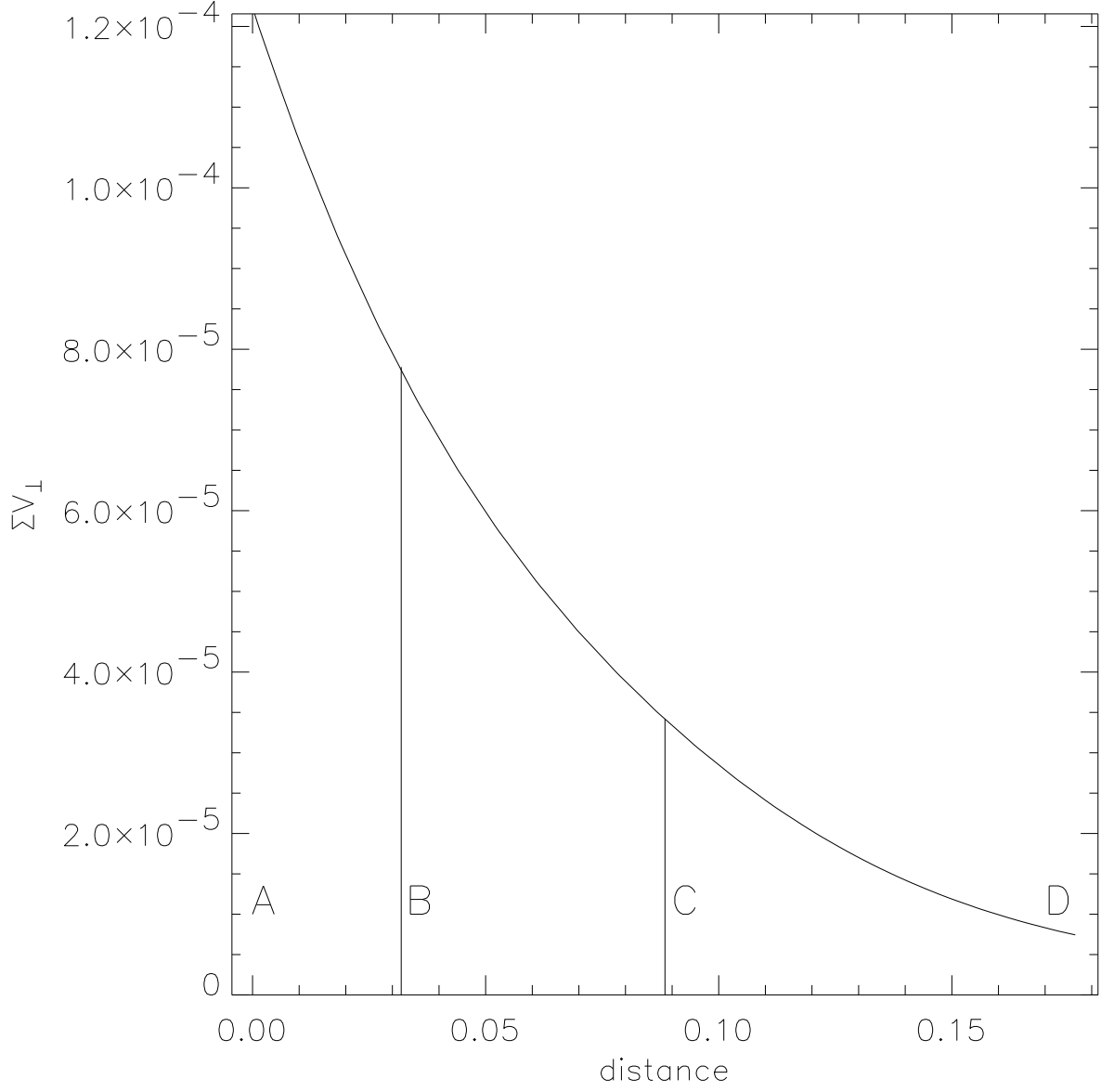}
\caption[fig6.eps]{Mass flux across dotted line AD
in Figure 4 as a function of distance along the line in units of 
$M_d \Omega_p/a$ within the disk gap.
Position A is in the outer disk region,
position B is on the streamline that becomes the outer boundary
of the L2 accretion stream, position C is on the streamline that becomes 
the inner boundary of the L2 accretion stream, position D is in the
horseshoe orbit region.
\label{fig6}}
\end{figure*}

\begin{figure*}
\figurenum{7}
\epsscale{1.0}
\plotone{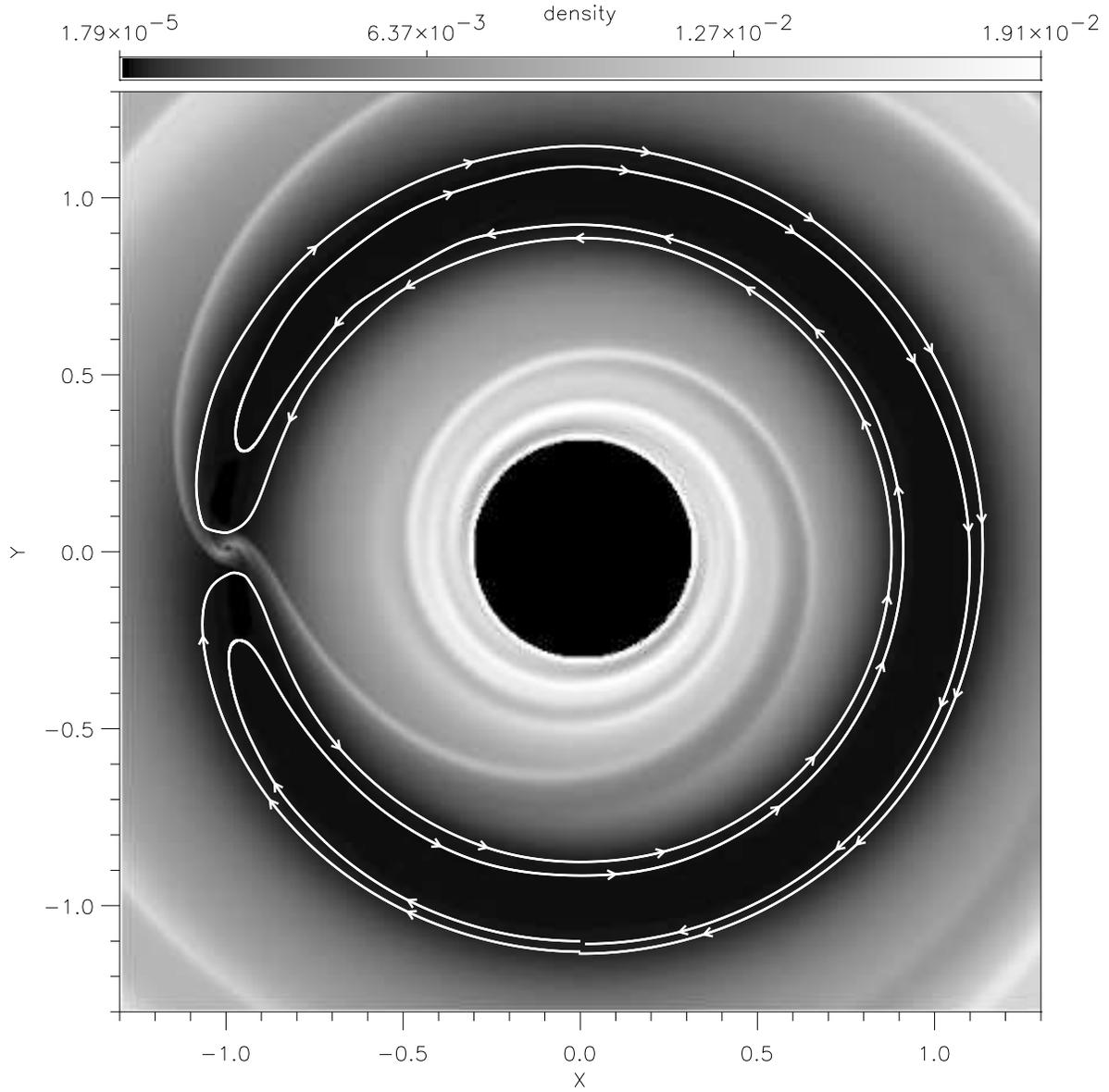}
\caption[fig7.eps]{
Two horseshoe orbits within the disk gap
(see section 4.3.3).
\label{fig7}}
\end{figure*}

\begin{figure*}
\figurenum{8}
\epsscale{1.0}
\plotone{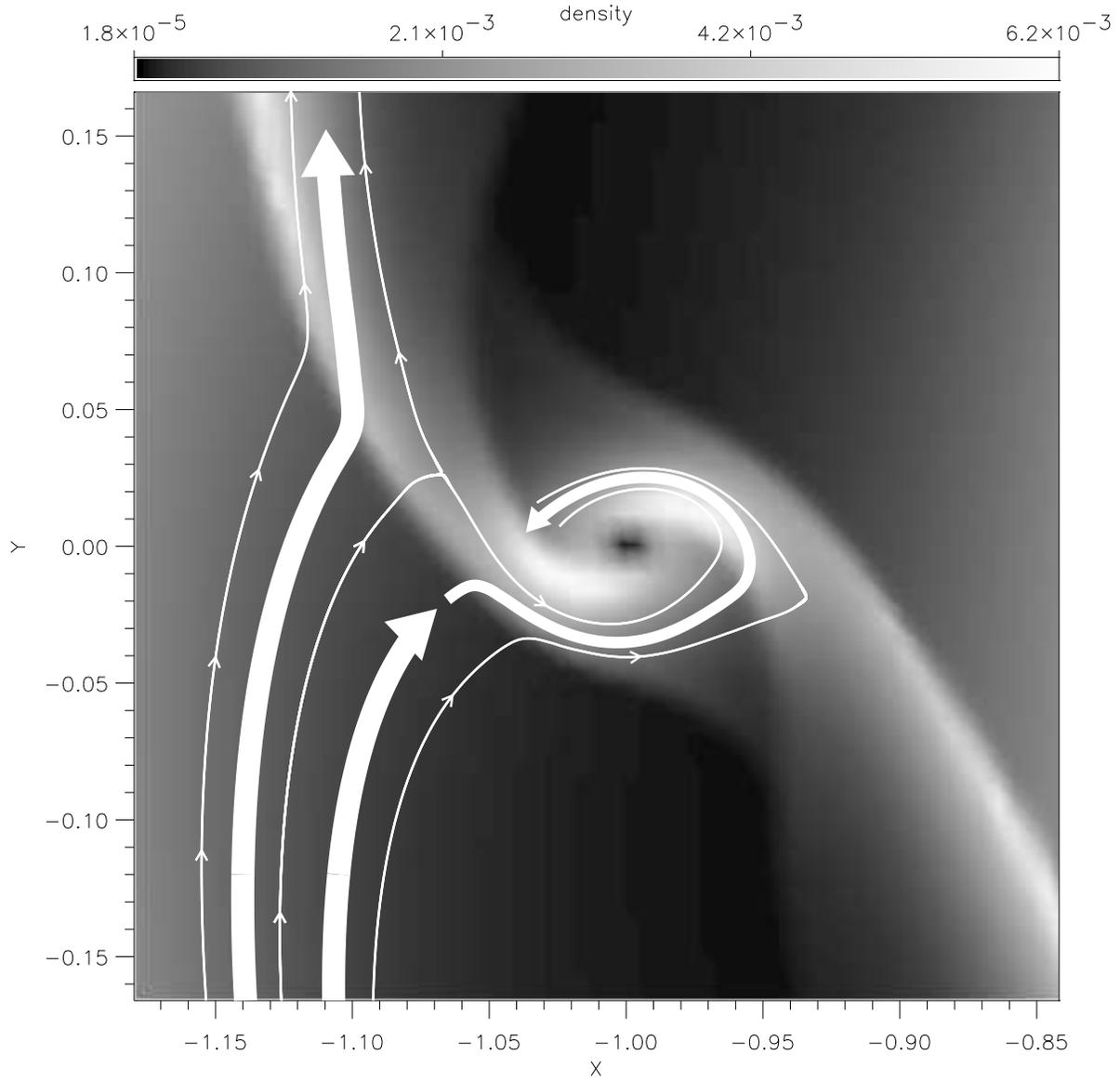}
\caption[fig8.eps]{
Flow evolution of gas stream material.
Material flowing along the left arrow 
orbits about the star, returns as material that flows
along the two right arrows, and becomes accreted by the planet
(see section 4.4).
\label{fig8}}
\end{figure*}

\begin{figure*}
\figurenum{9}
\epsscale{1.0}
\plotone{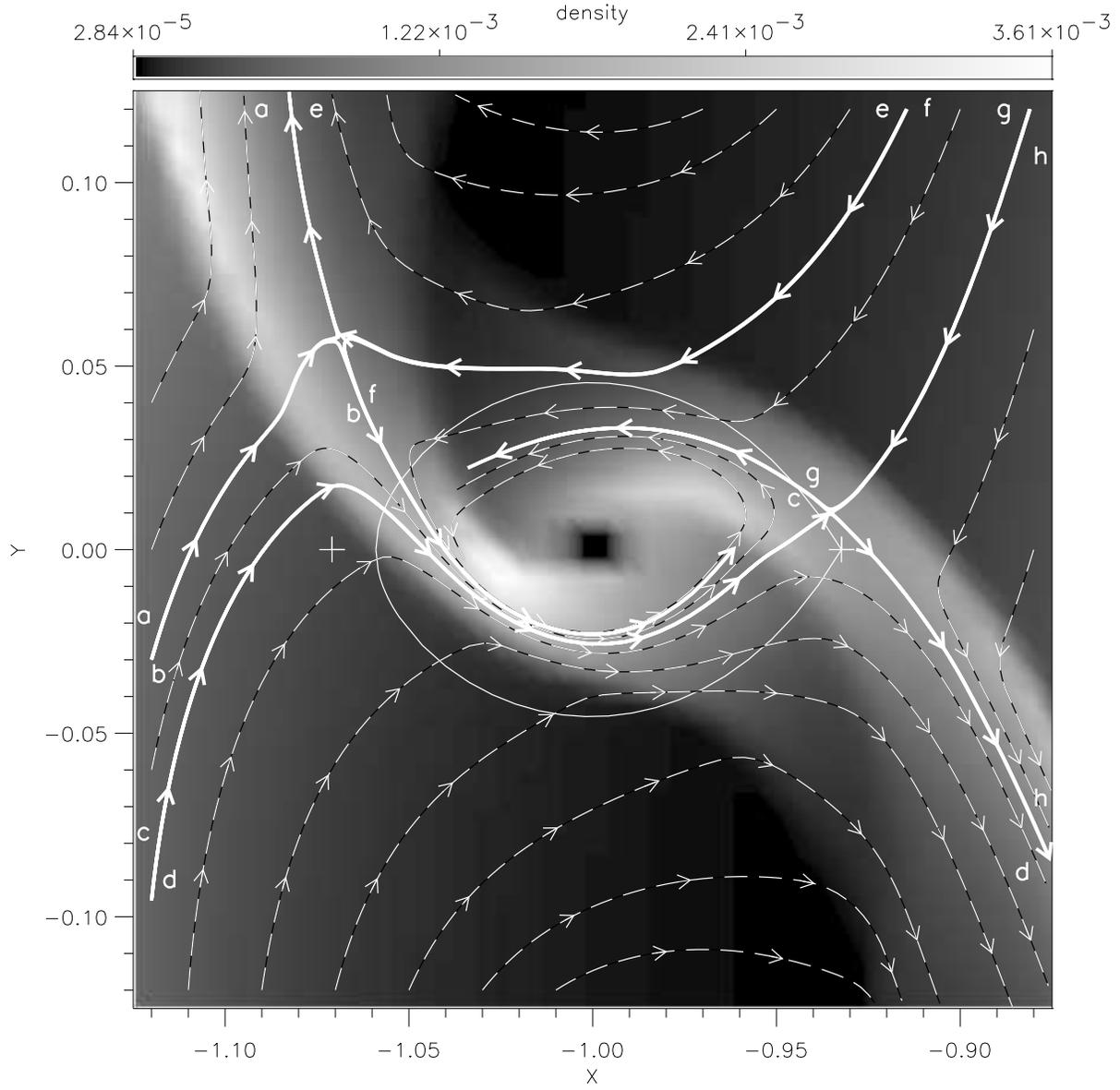}
\caption[fig9.eps]{High resolution image
of the vicinity of the Roche lobe.
Same as Figure 4, except that the simulation
was initialized to have a very small mass inner circumstellar
disk (see section 4.5).
\label{fig9}}
\end{figure*}

\begin{figure*}
\figurenum{10}
\epsscale{1.0}
\plotone{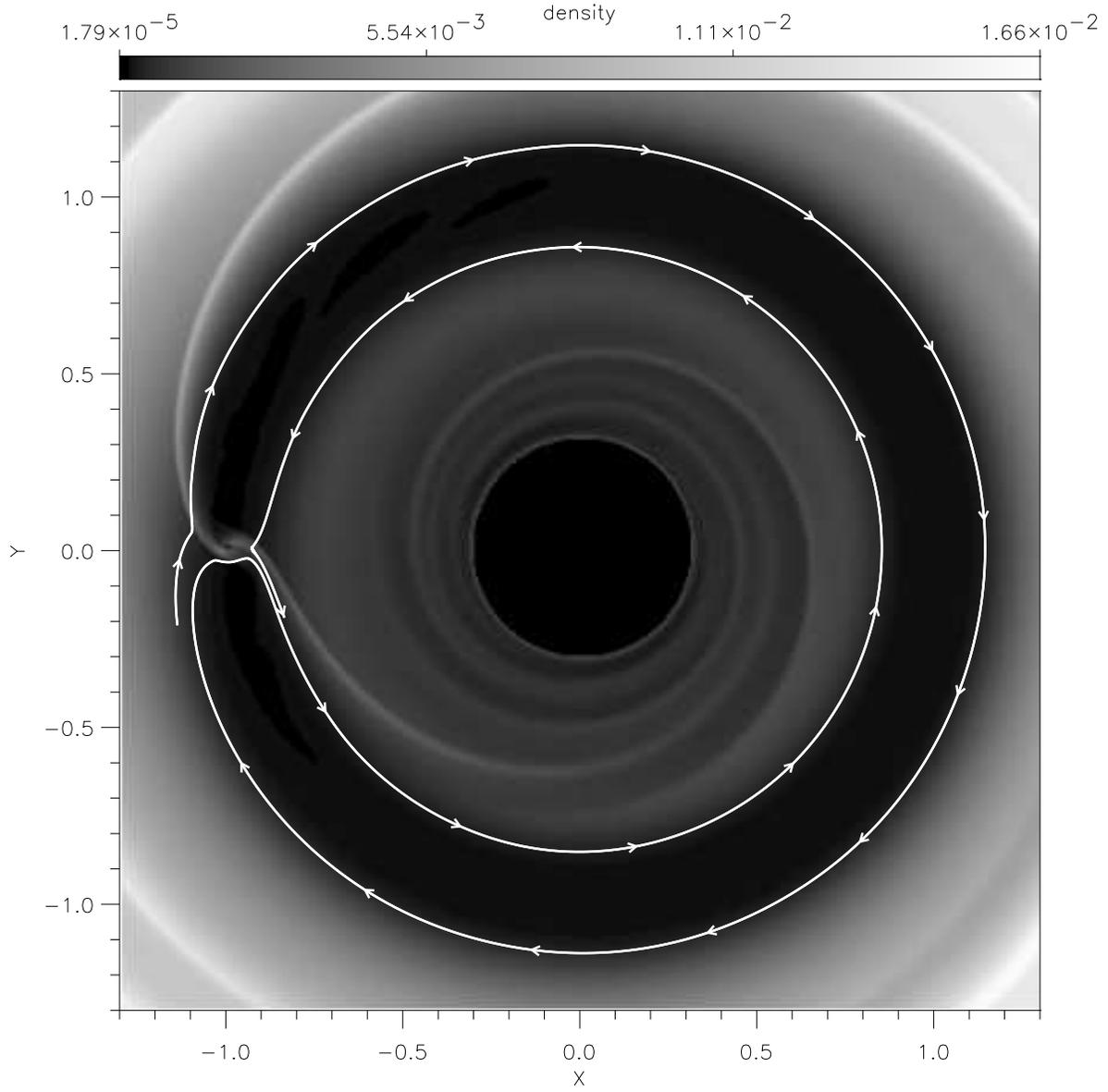}
\caption[fig10.eps]{A streamline
in the case of a depleted inner disk.
The streamline starts in the outer disk and
enters the inner disk
(see section 4.5).
\label{fig10}}
\end{figure*}

\begin{figure*}
\figurenum{11}
\epsscale{1.0}
\plotone{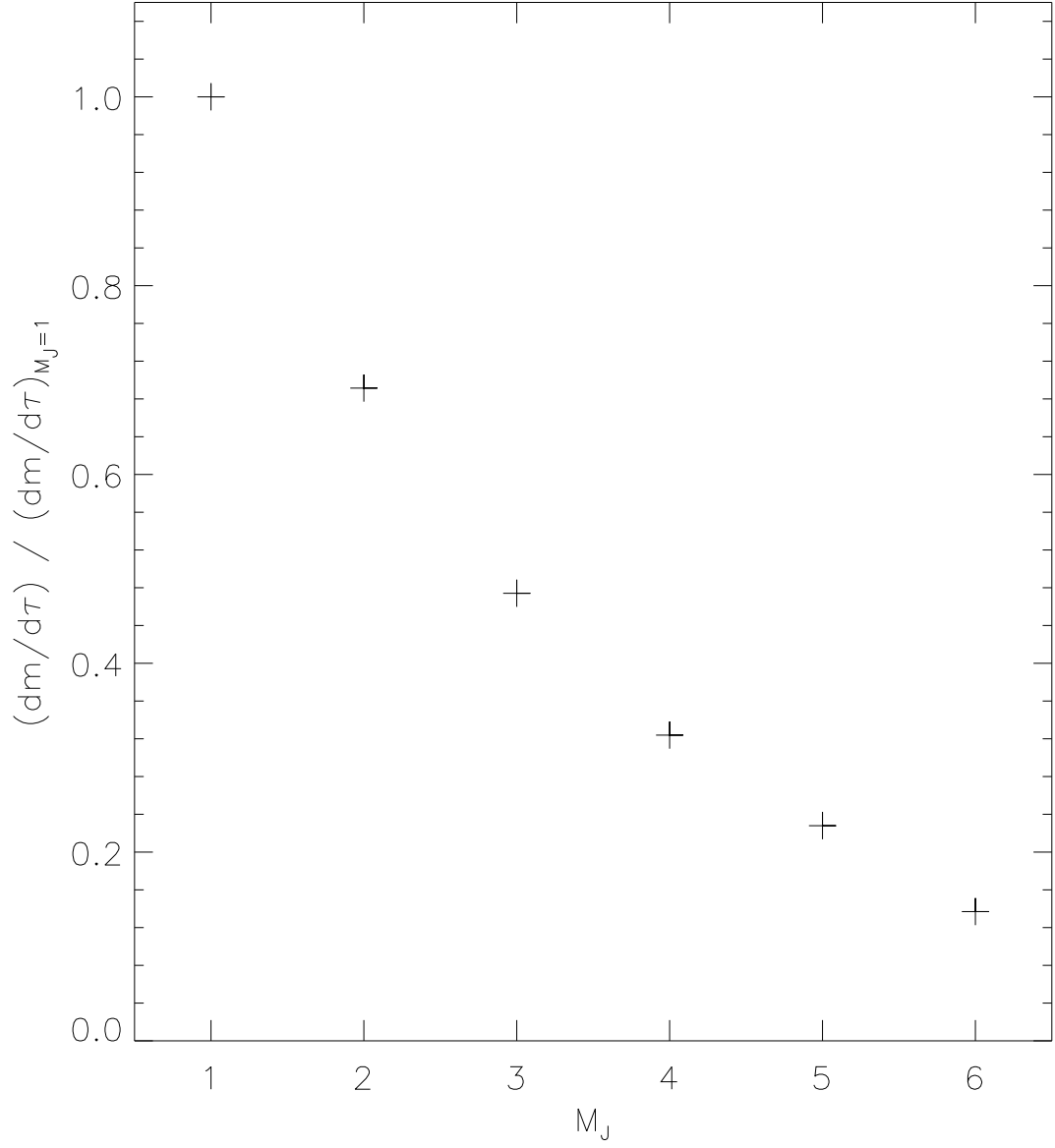}
\caption[fig11.eps]{Mass accretion rate
 (normalized by the mass accretion rate
onto a $1 M_J$ planet)  as a function
of planet mass for planets that orbit a $1 M_{\odot}$
star.
\label{fig11}}
\end{figure*}

\end{document}